\documentclass[12pt]{article}
\usepackage{amsmath,amssymb,amsfonts,epsf}
\usepackage[nosort]{cite}
\usepackage[margin=1in]{geometry}

\newcommand{\fft}[2]{{\frac{#1}{#2}}}
\newcommand{\ft}[2]{{\textstyle\frac{#1}{#2}}}

\def\nn{\nonumber}

\let\bm=\bibitem

\newcommand{\be}{\begin{equation}}
\newcommand{\ee}{\end{equation}}
\def\ba{\begin{array}}
\def\ea{\end{array}}
\def\ft#1#2{{\textstyle{\frac{\scriptstyle #1}{\scriptstyle #2}}}}
\def\fft#1#2{\frac{#1}{#2}}

\def\sst#1{{\scriptscriptstyle #1}}

\def\td{\tilde}

\def\dalemb#1#2{{\vbox{\hrule height .#2pt
        \hbox{\vrule width.#2pt height#1pt \kern#1pt
                \vrule width.#2pt}
        \hrule height.#2pt}}}

\newcommand{\bea}{\begin{eqnarray}}
\newcommand{\eea}{\end{eqnarray}}

\def\0{{\sst{(0)}}}
\def\1{{\sst{(1)}}}
\def\2{{\sst{(2)}}}
\def\3{{\sst{(3)}}}
\def\4{{\sst{(4)}}}
\def\5{{\sst{(5)}}}
\def\6{{\sst{(6)}}}
\def\7{{\sst{(7)}}}
\def\8{{\sst{(8)}}}

\def\ep{{\epsilon}}

\begin{document}

\begin{center}\ \\ \vspace{60pt}
{\Large {\bf Bound Orbits of Solar Sails and General Relativity}}\\ 
\vspace{30pt}
Roman Ya. Kezerashvili and Justin F. V\'azquez-Poritz
\vspace{20pt}

{\it Physics Department\\New York City College of Technology, The City University of New York\\ 300 Jay Street, Brooklyn NY 11201, USA.}\\
\vspace{20pt}
{\tt rkezerashvili@citytech.cuny.edu}\\ 
{\tt jvazquez-poritz@citytech.cuny.edu}

\end{center}

\vspace{30pt}

\centerline{\bf Abstract}

\noindent We study how the curvature of spacetime, in conjunction with solar radiation pressure (SRP), affects the bound orbital motion of solar sails. While neither the curvature of spacetime nor the SRP alter the form of Kepler's third law by themselves, their simultaneous effects lead to deviations from this law. We also study deviations from Keplerian motion due to frame dragging, the gravitational multipole moments of the sun, a possible net electric charge on the sun, and a positive cosmological constant. The presence of the SRP tends to increase these deviations by several orders of magnitude, possibly rendering some of them detectable. As for non-circular bound orbits, the SRP dampens the rate at which the perihelion is shifted due to curved spacetime, while the perihelion shift due to the oblateness of the sun is increased. With regards to the Lense-Thirring effect, the SRP increases the angle of precession of polar orbits during one orbital period, although the precession frequency is not actually altered. We also consider non-Keplerian orbits, which lie outside of the plane of the sun. In particular, we investigate how the pitch angle of the solar sail is affected by the partial absorption of light by the sail, general relativistic effects, and the oblateness of the sun. Non-Keplerian orbits exhibit an analog of the Lense-Thirring effect, in that the orbital plane precesses around the sun. A near-solar mission for observations of these effects could provide an interesting confirmation of these phenomena.

\vspace{0.1cm}

\thispagestyle{empty}

\pagebreak
\setcounter{page}{1}

\tableofcontents

\addtocontents{toc}{\protect\setcounter{tocdepth}{3}}

\newpage

\section{Introduction}

In the last decade, the observation and analysis of satellite motion has provided an abundance of data with which to test basic physical principles. Examples include the Pioneer anomaly, which is an unexplained acceleration of the Pioneer 10 and 11 spacecraft on escape trajectories from the outer solar system \cite{rk3,rk4}, and the flyby anomaly, for which the velocities of the Galileo, NEAR and Cassini spacecraft  are different from what is expected after Earth flybys \cite{AnderWilliams,Lammerzahal}. In fact, the difficulties of explaining these anomalies within the framework of standard physics became a motivation to speculate on the unlikely possibility that they originate from new physics. Missions have been proposed \cite{Lammerzahal,rk6,rk5} to further explore these anomalies, in order to better understand the laws of fundamental physics as they affect dynamics within the solar system. 

One of the most basic laws that describes motion in the solar system is Kepler's third law, which can be derived from Newton's law of gravitation and provides a relationship between the period $T$ and the orbital radius $r$ of an object orbiting the sun. Namely, $T^2$ is proportional to $r^3$ with the proportionality constant given by $4\pi ^{2}/GM$, where $G$ is the gravitational constant and $M$ is the mass of the sun. Below, we discuss deviations from Keplerian orbits due to phenomena within conventional physics, which can be observed from the motion of solar sail propelled (SSP) satellites \cite{rk1,rk2}.  

In the Newtonian approximation, the sun is the source of a gravitational force on other masses. In the general relativistic framework, in the absence of non-gravitational forces, objects follow geodesics on the curved spacetime in the vicinity of the sun. At the same time, the sun is also a source of solar electromagnetic radiation, which produces an external force on objects via the solar radiation pressure (SRP). It is of particular interest to analyze how the trajectory of an object deviates from a geodesic under the action of the force due to the solar radiation pressure. We will assume that the backreaction of the radiation on spacetime is negligible. Therefore, we can say that objects move in the {\it photo-gravitational} field of the sun. 

When an SSP satellite undergoes orbital motion within the photo-gravitational field of the sun, the orbital period is altered by the presence of the solar sail, since the force from the solar radiation pressure affects the dynamics of the orbit. Moreover, due to the continuously-acting solar radiation pressure, an SSP satellite is capable of exotic non-Keplerian orbits which are impossible for spacecraft which are not equipped with solar sails. The solar sail enables one to design trajectories for which the SSP satellite is effectively levitated above the sun-- namely, the sun lies outside of the orbital plane. For example, non-Keplerian orbits can be above the ecliptic plane and parallel to it.

Satellite orbits been extensively studied within the Newtonian approximation for gravity (see \cite{astrodyn}, for example), and many papers have considered general relativistic effects on orbits. However, general relativistic effects on solar sails, which are propelled by electromagnetic radiation pressure, has largely remained an unexplored subject. At first glance, such an intersection of topics might not appear necessary. However, we currently have the technology to put solar sails in orbit around the sun with heliocentric distances approaching $0.05$ AU. At such a proximity to the sun, which is considerably closer than the orbit of Mercury, the continuous effects of curved spacetime on bound orbits should be considered. 

We recently considered various general relativistic effects on solar sails in circular orbits within the plane of the sun \cite{poritz}. The curvature of spacetime can be separated into a contribution associated with a static central mass and a portion corresponding to the rotation of the sun. For the purpose of comparison, we also considered a number of other effects, such as the gravitational multipole moments of the sun, a possible small net electric charge on the sun, and a small positive cosmological constant. 

In this paper, we generalize this study to include slightly non-circular orbits, orbits at arbitrary polar angle (since spherical symmetry is broken by frame dragging effects), and non-Keplerian orbits. We find that the perihelion shift of non-circular orbits which is due to curved spacetime in general relativity actually occurs at a slower rate due to the SRP. On the other hand, there is also a perihelion shift that occurs in the reverse direction due to the oblateness of the sun. This second perihelion shift is also present within Newtonian gravity, and we find this effect to be augmented by the SRP. We also consider the Lense-Thirring effect, which is the precession of polar orbits due to frame dragging, and which has no counterpart in Newtonian gravity. Perhaps surprisingly, the precession frequency is not altered by the SRP. We also consider various effects on solar sails undergoing non-Keplerian orbits, such as the partial absorption of sunlight by the surface of the sail, spacetime curvature and frame dragging, and the oblateness of the sun. Interestingly enough, we find an analog of the Lense-Thirring effect for non-Keplerian orbits, for which the orbital plane (which lies outside of the plane of the sun) precesses around the sun.

This paper is organized as follows. In section 2, we consider various effects on bound orbits in the plane of the sun, including solar radiation pressure, static curved spacetime, frame dragging, the gravitational multipole moments of the sun, a possible small net electric charge on the sun, and a small positive cosmological constant. In section 3, we consider circular orbits out of the plane of the sun. After a brief review of non-Keplerian orbits for a perfectly reflecting solar sail, we consider the effects of light absorption, spacetime curvature and frame dragging, and the oblateness of the sun. Conclusions follow in section 4, where we summarize our results, and consider a solar-sail propelled satellite with specified structural and orbital parameters in order to do a comparative analysis on the how the SRP affects various phenomena.

\section{Bound orbits in the plane of the sun}

The purpose of this section is to point out various sources of deviations from Keplerian orbits and discuss how the resulting change in period is enhanced by the SRP to the degree in which it may be observed for some cases. The phenomena discussed include the curvature of spacetime in the vicinity of the sun, described by the Schwarzschild metric, frame dragging due to the rotation of the sun, for which the curved spacetime is described approximately by the large-distance limit of the Kerr metric \cite{kerr}, the oblateness of the sun, the effect of a possible small net electric charge on  the sun, and a small positive cosmological constant. This list is by no means exhaustive. Some effects that we do not consider are the perturbations in motion due to the gravitational fields of the planets, the magnetic field of the sun and the solar wind. While the reality is that all effects occur simultaneously, particular effects may be isolated by considering a variety of orbits. However, in order to get a fairly accurate idea of the {\it relative} importance of some effects in the presence of the SRP, we will consider them individually for the simple scenario of circular orbits.

\subsection{Solar radiation pressure}

We will be mainly considering an SSP satellite orbiting the sun. According to Maxwell's electromagnetic theory, electromagnetic waves carry the energy and linear momentum and the radiation
pressure $P$ exerted on the surface of satellite or solar sail due to
momentum transport by photons is given by
\be
P=\fft{2\eta S}{c}\,,\qquad S=\frac{L_S}{4\pi r^{2}}\,,
\ee
where $S$ is the magnitude of the Poynting vector and the solar luminosity is $L_S=3.842\times 10^{26}$ W. Also, $0.5\le \eta\le 1$, where $\eta=0.5$ corresponds to the total absorption of photons by the surface of the satellite and $\eta=1$ corresponds to the total reflection of the solar radiation. The resulting force on the SSP satellite is $F=PA$, where $A$ is the area of solar sail directly facing the sun. Thus, the acceleration due to this force can be expressed as
\be\label{kappa}
a=\fft{\kappa}{r^2}\,,\qquad \kappa\equiv \frac{\eta L_S}{2\pi c\sigma}\,,\qquad \sigma =\frac{m}{A}\,,
\ee
where $m$ is the mass of the SSP satellite. The mass per area $\sigma$ is a key design parameter for solar sails \cite{rk1,rk11,rk12}. 

\begin{figure}[ht]
   \epsfxsize=2.6in \centerline{\epsffile{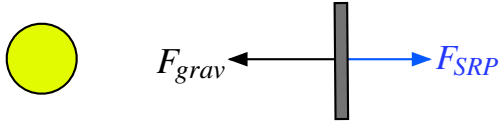}}
   \caption[FIG. \arabic{figure}.]{\footnotesize{The gravitational force and solar radiation pressure force acting on a solar sail which is directly facing the sun.}}
\label{fig3}
\end{figure}

We will first consider the effect of the SRP on Keplerian orbits in the Newtonian approximation for gravity. The SRP force is repulsive and the gravitational force is attractive. For simplicity, we will start by restricting ourselves to the case in which the surface of the solar sail is directly facing the sun, so that vectors normal to the sail are directed along the sun-satellite line. Then both forces act along the same line, as shown in Figure \ref{fig3}. 
Also, both forces fall off as $1/r^2$, since $r$ is the heliocentric distance. Therefore, the consideration of both forces leads to a modification of the effective mass of the sun in Kepler's third law. Namely, the mass of the sun, which is $M=1.99\times 10^{30}$ kg, can be effectively renormalized as $\td M\equiv M-\kappa/G$. 

The modified Kepler's third law can be expressed as
\be\label{thirdlaw}
T^2=\fft{4\pi^2}{G\td M}\ r^3\,,
\ee
where $r$ is the radius in the case of circular orbits, while for elliptical orbits we replace $r$ by the length $a$ of the semi-major axis of the ellipse. Since $\td M<M$, the orbital period of an SSP satellite will always be longer than that of a conventional satellite for a given orbital radius. In the Newtonian approximation, Kepler's third law retains its form in the presence of the SRP. Eq. (\ref{thirdlaw}) can easily be obtained from the usual expression for Keplerian orbits by simply replacing the solar mass $M$ with the effective solar mass $\td M$. This is just a reflection of the fact that both the Newtonian force of gravity and the solar radiation pressure force go as the inverse square of the heliocentric distance. One can show that the orbital radius $r$ (length of the semi-major axis $a$ for elliptical orbits) can be expressed in terms of the energy of the SSP satellite. Thus, the period can be expressed in terms of the energy as well as the solar sail parameter $\kappa$. We can also use (\ref{thirdlaw}) to express $\kappa$ in terms of the period and radius as
\be\label{k1}
\kappa=GM-\fft{4\pi^2r^3}{T^2}\,,
\ee
which is useful in the design of SSP satellites.

As a first example, we will consider Mercury, whose average orbital radius is $r=5.79\times 10^{10}$ m,  which corresponds to a period of about $87.9686$ days. Mercury has a mass of $3.30\times 10^{23}$ kg and a radius $r_M=2.44\times 10^6$ m. This yields $\sigma=1.76\times 10^{10}$ kg/m$^2$, where the effective area is $\pi r_M^2$. As a first approximation, we assume that no sunlight is reflected by Mercury, so that $\eta=0.5$. This yields an increase in period on the order of $10^{-7}$ s which, as to be expected for any planet, is negligible. 

We will now consider a conventional satellite orbiting the sun at a distance of $r=1$ AU$\approx 1.50\times 10^{11}$ m, which corresponds to a period of one year. If the mass of the satellite is $1000$ kg and its area is $2$ m$^2$, then $\sigma=500$ kg/m$^2$. If we suppose that $\eta=0.75$, then the increase in period due to SRP is about $36$ s, which could be observed.

In the remainder of this section, we will consider an SSP satellite with the following specifications:
\bea\label{values}
r &=& 0.05\ \mbox{AU}\approx 7.48\times 10^9\ \mbox{m}\,,\nn\\
\eta &=& 0.85\,,\quad \sigma=0.00131\ \mbox{kg/m}^2\,.
\eea
If the acceleration due to the SRP is ignored, then the corresponding orbital period would be about $4$ days. When the SRP is taken into account, the period is about $70$ days. 

Since we are considering motion in a plane, we can use polar coordinates $(r,\phi)$. We can derive the orbital equation from the following expressions for conserved quantities
\bea
E_N &=& \fft12 \dot r^2+\fft12 r^2 \dot\phi^2-\fft{G\td M}{r}\,,\nn\\
L_N &=& r^2\dot\phi\,,
\eea
where dots denote time derivatives, and $E_N$ and $L_N$ are the energy and angular momentum per unit mass of solar sail, respectively. The subscripts $N$ are used to distinguish these conserved quantities from their general relativistic counterparts. Then the orbital equation is given by
\be\label{Newtonorbital}
\left( \fft{dr}{d\phi}\right)^2 = \left(2E_N+\fft{2G\td M}{r}-\fft{L_N^2}{r^2}\right) \fft{r^4}{L_N^2}\,,
\ee
and a solution is
\be
r^{-1}=\fft{G\td M}{L_N^2}+\fft{\sqrt{2E_N+G^2\td M^2/L_N^2}}{L_N}\ \cos \phi\,.
\ee

\subsection{Static curved spacetime}

We will now consider the simultaneous effects of the SRP and curved spacetime in the vicinity of the sun. We assume that the backreaction of the electromagnetic radiation on the background geometry is negligible so that it acts on the SSP satellite only via the SRP. The exterior spacetime of the sun is approximately described by the Schwarzschild metric, which is given by
\be\label{form}
ds^2=-f c^2dt^2+f^{-1} dr^2+r^2 d\Omega^2\,,
\ee
where the metric function $f$ is given by
\be\label{f}
f=1-\fft{2GM}{c^2r}\,.
\ee
Also, $d \Omega^2$ refers to the metric of a unit two-sphere $S^2$, which can be written in terms of the polar angle $\theta$ and azimuthal angle $\phi$ as
\be
d\Omega^2=d\theta^2+\sin^2\theta\ d\phi^2\,.
\ee
Spherical symmetry allows us to orient the coordinate system so that the orbit is confined to the equatorial plane at $\theta=\pi/2$, and thus $p_{\theta}=0$. Since the metric is independent of time and the azimuthal direction $\phi$, the corresponding components $p_t$ and $p_{\phi}$ of the 4-momentum are conserved. We define the constants of motion $E\equiv -p_t/m$ and $L\equiv p_{\phi}/m$, where $m$ is the rest mass of the SSP satellite. Thus,
\be\label{p1}
p^t =\fft{m E}{c^2f},\quad p^r=m\ \fft{dr}{d\tau},\quad p^{\theta}=0,\quad p^{\phi}=\fft{m}{r^2}L,
\ee
where $\tau$ is the proper time. In the absence of the SRP, $p^2=-m^2c^2$ yields
\be\label{reqn1}
\left( \fft{dr}{d\tau}\right)^2=\fft{E^2}{c^2}-\left(c^2+\fft{L^2}{r^2}\right) f\,.
\ee
Differentiation of (\ref{reqn1}) with respect to $\tau$ gives the radial component of the 4-acceleration
\be\label{firstar}
a^r=\fft{d^2r}{d\tau^2}+\fft{GM}{r^2}-\fft{L^2}{r^3}+\fft{3GM L^2}{c^2r^4}\,.
\ee

We will now turn on the SRP, so that
\be\label{ar}
a^r=\fft{\kappa}{r^2}\,,
\ee
where $\kappa$ is given in (\ref{kappa}). Note that even though the coordinate $r$ does not measure the proper distance, the surface area of a sphere is still given by $4\pi r^2$, which means that the magnitude of the Poynting vector as well as the acceleration are given by the same expressions as in the Newtonian approximation. Equating the expressions for $a^r$ given in (\ref{firstar}) and (\ref{ar}) and taking the first integral gives
\be\label{sailreqn}
\left( \fft{dr}{d\tau}\right)^2=\fft{E^2}{c^2}-c^2+\fft{2G\td M}{r}-\fft{L^2}{r^2}f\,.
\ee
From (\ref{sailreqn}) and the $\phi$ equation in (\ref{p1}), we find the orbital equation
\be\label{curvedorbital}
\left( \fft{dr}{d\phi}\right)^2=\left(\fft{E^2}{c^2}-c^2+\fft{2G\td M}{r}-\fft{L^2}{r^2}f\right) \fft{r^4}{L^2}\,.
\ee
Note that the SRP reduces the effective mass only in the term which is present for Newtonian gravity. In the limit of Newtonian gravity and non-relativistic speeds, $E^2/c^2-c^2\rightarrow 2E_N$ and the orbital equation (\ref{curvedorbital}) reduces to (\ref{Newtonorbital}).

\subsubsection{Circular orbits}

We will first consider orbits which are completely circular, for which
\be\label{circular}
\fft{dr}{d\tau}=0\,,\qquad \fft{d^2r}{d\tau^2}=0\,.
\ee
This yields
\be
E^2 = c^4+\left( \fft{4GM-c^2r}{c^2r-3GM}\right) \fft{c^2G\td M}{r},\quad
L^2 = \fft{c^2 G\td M r^2}{c^2r-3GM}.
\ee
The above equation for $L^2$ can be inverted to get
\be\label{radius}
r=\fft{L^2+ \sqrt{L^4-12G^2 L^2 M\td M/c^2}}{2G\td M}\,.
\ee
We can write (\ref{sailreqn}) in terms of an effective potential as
\be
c^2\left( \fft{dr}{d\tau}\right)^2=E^2-V_{eff}^2(r)\,,
\ee
where 
\be
V_{eff}=\sqrt{c^2-\fft{c^2G\td M}{r}\left( \fft{4GM-c^2r}{c^2r-3GM}\right)}\,.
\ee
The minimum of $V_{eff}$ is at the radius given by (\ref{radius}), which therefore corresponds to a stable circular orbit. 

Using $dt/d\tau=p^t/m$ and $d\phi/d\tau=p^{\phi}/m$, we find for the orbital period $T$:
\be\label{period}
T^2=\fft{4\pi^2r^3}{G\td M} \left[ 1+\kappa\ \fft{c^2r-4GM}{(c^2r-2GM)^2}\right]\,.
\ee
If we consider the case of $\kappa =0$ in (\ref{period}), which means that there is no effect from SRP, then we obtain the classical Keplerian expression for the orbital period. Therefore, the description of the motion of an object in the static exterior spacetime of the sun described by the Schwarzschild metric does not alter Keplerian circular orbits. While the solar radiation pressure diminishes the effective solar mass, the form of Kepler's third law remains intact. It is only through the simultaneous effects of general relativity, via the Schwarzschild metric, and solar radiation pressure on the orbital motion of the SSP satellite gives the deviation from Kepler's third law by the radial dependent (and thus energy dependent)  factor in the squared brackets. 

Keeping only the leading correction due to the curvature of spacetime, (\ref{period}) reduces to
\be\label{curvedT}
T^2\approx\fft{4\pi^2r^3}{G\td M} \left[ 1+\fft{\kappa}{c^2r}\right]\,.
\ee
For the specification given in (\ref{values}), we find that this yields an increase in the period of about $0.6$ s. 

Inverting (\ref{curvedT}) to express the $\kappa$ parameter in terms of the period and radius gives
\be
\kappa\approx\fft{GMT^2-4\pi^2r^3}{T^2+4\pi^2r^2/c^2}\,.
\ee

\subsubsection{Non-circular orbits}

We will now calculate the precession of orbits which are not completely circular. We define the coordinate
\be\label{y}
y=\fft{1}{r}-\fft{G\td M}{L^2}\,.
\ee
For a nearly circular orbit, $y$ is small. Thus, neglecting terms in $y^3$, the orbital equation (\ref{curvedorbital}) can be expressed as
\be
\left( \fft{dy}{d\phi}\right)^2=\fft{E^2/c^2+G^2\td M^2/L^2-c^2}{L^2}+\fft{2G^4M\td M^3}{c^2 L^6}+\fft{6G^3M\td M^2}{c^2L^4}y+\left( \fft{6G^2M\td M}{c^2 L^2}-1\right) y^2\,.
\ee
The solution is
\be\label{precsol}
y=y_0+A \cos [k(\phi -\phi_0)]\,,
\ee
where $\phi_0$ is an arbitrary constant and
\bea
k &=& \sqrt{1-\fft{6G^2 M\td M}{c^2 L^2}}\,,\nn\\
y_0 &=& \fft{3G^3 M\td M^2}{c^2 L^4 k^2}\,,\nn\\
A &=& \fft{1}{k} \sqrt{\fft{E^2/c^2+G^2 \td M^2/L^2-c^2}{L^2}+\fft{2GM\td M^3}{c^2 L^6}+\fft{6G^3 M\td M^2}{c^2 L^4}y_0-k^2y_0^2}\,.
\eea
Note that the orbit oscillates about $y=y_0$ which corresponds to the radius for a circular orbit in general relativity, whereas $y=0$ corresponds to the radius for a circular orbit in Newtonian gravity. The perihelion shift during one complete orbit is
\be\label{pshift}
\Delta\phi=\fft{2\pi}{k}-2\pi\,.
\ee
For nearly Newtonian orbits,
\be
\Delta\phi\approx \fft{6\pi GM}{c^2 r}\,,
\ee
where we have taken $L^2\approx G\td M r$. Note that the perihelion shift per orbit is not affected by the SRP at this approximation. In fact, this is exactly the formula which describes the perihelion shift for the orbit of Mercury, which is about $0.43$ arcseconds per year. The perihelion shift of a conventional satellite at $r=0.05$ AU is substantially larger-- about $70$ arcseconds per year. Interestingly enough, this is one scenario for which the SRP would dampen the effect, since it increases the period of the orbit, and therefore the time required for the perihelion shift $\Delta\phi$ to occur. Using the approximate period given by (\ref{thirdlaw}), we find that the rate at which the perihelion is shifted is
\be
\fft{\Delta\phi}{T}\approx \fft{3GM\sqrt{G\td M}}{c^2r^{5/2}}\,.
\ee
As an example, for our SSP satellite the perihelion shift is reduced to about $4$ arcseconds per year.

\subsection{Frame dragging}

The rotation of the sun causes frame dragging, which affects the trajectories of orbiting objects. 
The external spacetime of a slowly rotating body with mass $M$ and angular momentum $J$ is described approximately by the large-distance limit of the Kerr metric \cite{kerr}
\be\label{kerrmetric}
ds^2=-f c^2 dt^2-\fft{4GJ}{c^2r}\sin^2\theta\ dt d\phi+\fft{dr^2}{f}+r^2 d\Omega^2,
\ee
where $f$ is given by (\ref{f}). We do not use the full Kerr metric since it does not seem to describe the external spacetime of a rotating material body, because it does not smoothly fit onto metrics which describe the interior region occupied by physical matter. Since there are corrections to this metric in higher-order $J$ we will work up to only linear order in $J$, which suffices for the slowly rotating sun. 
Note that $J>0$ for a prograde orbit with respect to the sun, while $J<0$ for a retrograde orbit. 

As previously, we define the constant of motion $E\equiv -p_t/m$. The spherical symmetry of the Schwarzschild metric enabled us to orient orbits to lie within the equatorial plane. However, we no longer have this luxury for the metric (\ref{kerrmetric}). Thus, there is now a constant of motion associated with the component of angular momentum per unit mass that is normal to the equatorial plane, and an additional constant of motion associated with the total angular momentum per unit mass. Namely, $L_z\sin\theta\equiv p_{\phi}/m$, where the $z$ subscript denotes the direction that is normal to the equatorial plane, and
\be\label{ptheta}
p^{\theta}=\fft{m}{r^2} \sqrt{P-\fft{L_z^2}{\tan^2\theta}}\,,
\ee
where the Carter integral $P$ is related to the total angular momentum \cite{carter,bardeen}. Note that $P=0$ corresponds to motion in the equatorial plane, for which $\theta=\pi/2$ and $p^{\theta}=0$. No other orbits can lie within a fixed plane.

It will be useful to have the contravariant versions of $p_t$ and $p_{\phi}$. Inverting the $(t,\phi)$ portion of the metric (\ref{kerrmetric}) gives
\be
g^{tt}=-\fft{1}{c^2f}\,,\qquad g^{t\phi}=-\fft{2GJ}{c^4fr^3}\,,\qquad g^{\phi\phi}=\fft{1}{r^2\sin^2\theta}\,,
\ee
which leads to
\bea\label{pt}
p^t &=& g^{tt} p_t+g^{t\phi}p_{\phi}=\fft{mE}{c^2f}-\fft{2GmJL_z\sin\theta}{c^4fr^3}\,,\nn\\
p^{\phi} &=& g^{\phi\phi}p_{\phi}+g^{t\phi}p_t=\fft{mL_z}{r^2\sin\theta}+\fft{2GmJE}{c^4r^3f}\,.
\eea

In the absence of the solar radiation pressure, $p^2=-c^2m^2$ yields
\be\label{reqnkerr}
\left( \fft{dr}{d\tau}\right)^2=\fft{E^2}{c^2}-f\left( c^2+\fft{L_z^2+P}{r^2}\right) -\fft{4GJEL_z}{c^4r^3}\,.
\ee
Differentiating this with respect to $\tau$ gives
\be\label{arkerr}
a^r= \fft{d^2r}{d\tau^2}+\fft{GM}{r^2}-\fft{L_z^2+P}{r^3}+\fft{3Gc^2M(L_z^2+P)-6GJEL_z}{c^4r^4}\,.
\ee
Turning on the solar radiation pressure, the equation of motion is given by (\ref{ar}), which is not altered by linear terms in $J$. Taking the first integral  of $a^r$ yields
\be
\left( \fft{dr}{d\tau}\right)^2= \fft{E^2}{c^2}-c^2+\fft{2G\td M}{r}-\fft{f(L_z^2+P)}{r^2} -\fft{4GJEL_z}{c^4r^3}\,.
\ee

For a circular orbit, $dr/d\tau=0$ and $d^2r/d\tau^2=0$. Thus, from (\ref{reqnkerr}) and (\ref{arkerr})
\bea\label{ELz}
E &=& c\sqrt{\fft{X}{r(c^2r-3GM)}} - \fft{cJ}{r^2} \sqrt{\fft{G^2Y}{(c^2r-3GM)^3}}\,,\nn\\
L_z &=& c\sqrt{\fft{Y}{c^2r-3GM}} - \fft{3GJ}{c} \sqrt{\fft{X}{r(c^2r-3GM)^3}}\,.
\eea
where
\bea
X &\equiv& c^4r^2-c^2G(3M+\td M)r+4G^2M\td M\,,\nn\\
Y &\equiv& r(G\td M r-P)+\fft{3GMP}{c^2}\,.
\eea
We will now consider the special cases of orbits in the equatorial plane of the sun and polar orbits separately.

\subsubsection{Orbits in the equatorial plane of the sun}

For rotation in the equatorial plane ($\theta=\pi/2$ and $P=0$), from $dt/d\tau=p^t/m$ and $d\phi/d\tau=p^{\phi}/m$, we find the orbital period to be given by
\be\label{Tkerr}
T^2 \approx \fft{4\pi^2 r^3}{G\td M} \left[ 1+\kappa \fft{(c^2r-4GM)}{(c^2r-2GM)^2}\right] \left[ 1+\fft{2\sqrt{G} J\Big( 1+\fft{4\kappa (c^2r-GM)}{(c^2r-2GM)^2}\Big)}{c^2\sqrt{\td M} r^{3/2} \sqrt{1+\fft{\kappa (c^2r-4GM)}{(c^2r-2GM)^2}}}\right]\,.
\ee
We have presented $T^2$ as a product of three factors. The first factor is the same as in the Newtonian approximation for gravity in the presence of the SRP, and embodies the fact that the SRP effectively renormalizes the solar mass. The second factor shows the deviation from Keplerian orbits due to the simultaneous effects of SRP and the curvature of spacetime outside of a static central body. The third factor gives the deviation from Keplerian orbits from the combined effects of the SRP and frame dragging due to the rotation of the sun. Note that the second and third factors in (\ref{Tkerr}) involve both the solar mass $M$ and the parameter $\kappa$ separately, rather than simply in the combination of the renormalized solar mass $\td M$. For the case of $\kappa=0$, which means that there are no SRP effects, (\ref{Tkerr}) reduces to
\be
T^2=\fft{4\pi^2r^3}{G M}\left[ 1+\fft{2\sqrt{G}J}{c^2 \sqrt{M}r^{3/2}}\right]\,.
\ee
Thus, in the absence of the SRP there is still a deviation from Keplerian orbits due to frame dragging.

Keeping only the leading contributions due to spacetime curvature in (\ref{Tkerr}) gives
\be
T^2 \approx \fft{4\pi^2 r^3}{G\td M} \left[ 1+\fft{\kappa}{c^2r}\right] \left[ 1+\fft{2\sqrt{G} J}{c^2\sqrt{\td M} r^{3/2}}\right]\,.
\ee
Note that the SRP increases the effect that frame dragging has on the orbital period, since it is the renormalized mass $\td M$ that appears in the last factor. The speed of the outer layer of the sun at its equator is $v\approx 2000$ m/s at the equatorial radius $R\approx 7\times 10^8$ m, and we will assume that the core of the sun rotates with the same angular speed. Therefore, $J=\ft25 Mv R\approx 10^{42}$ kg m$^2$/s. Without the effects of the SRP, frame dragging leads to an increase (decrease) in the period of only $4\times 10^{-5}$ s for a prograde (retrograde) orbit. With the SRP and the specifications in (\ref{values}), frame dragging leads to a change in the period of about $0.01$ s.

\subsubsection{Polar orbits}

We will now consider orbital planes which pass through the poles of the sun. For this case, $L_z=0$ and the angle along the orbit is $\theta$. Setting $L_z=0$ in (\ref{ELz}) gives
\be
P=\fft{c^2G\td Mr^2}{c^2r-3GM}\,.
\ee
Note that the rotation of the sun does not effect this expression up to linear order in $J$. From the expressions for $p^t$ and $p^{\theta}$ in (\ref{pt}) and (\ref{ptheta}), we find the period of a polar orbit to be given by (\ref{period}). That is, up to linear order in $J$, the period of polar orbits is not affected by the effects of frame dragging.

However, the orbital plane of a non-equatorial orbit will precess in  the $\phi$ direction, which is a well-known effect of frame dragging called the Lense-Thirring effect. From (\ref{pt}), we find that the precession frequency for a polar orbit is
\be\label{precession}
\omega_p=\fft{d\phi}{dt}=\fft{2GJ}{c^2 r^3}\,.
\ee
Up to linear order in $J$, the precession frequency is not altered by the SRP. In addition, the angular speed given by (\ref{precession}) coincides with that of a particle falling from infinity with zero angular momentum. Thus, the angular speed in the $\phi$ direction is not altered by the polar orbital motion. Also, the precession frequency does not depend on the location of the satellite in the polar direction, although this would not be the case if quadratic terms in $J$ were considered. We will use these facts to find the approximate precession frequency for non-Keplerian orbits in section 3.

The angle of precession during one orbital period is given by
\be
\Delta\phi\approx \fft{4\pi GJ}{c^2 \sqrt{G\td M r^3}}\,,
\ee
which is increased by the SRP due to the corresponding increase in period. For both a standard and SRP satellite in a polar orbit at $r=0.05$ AU, we find the rate of precession to be about $0.03$ arcseconds per year.

\subsection{Gravitational multipole moments of the sun}

We will now consider the effect of the mass multipoles of the sun on bound orbits. In particular, the dominant higher moment is the quadrupole, which is associated with the oblateness of the sun. Note that we are still taking the sun to be a point-like light source. Our assumption is that variations in the SRP associated with the shape of the sun are negligible compared to the gravitational effects of oblateness. 

\subsubsection{Circular orbits}

Working in Newtonian gravity, the external gravitational potential of an oblate spheroid is given by \cite{oblateness}
\be\label{oblateV}
V=-\fft{G\td M}{r}-\fft{GM}{r}\sum_{n=2}^{\infty} J_n \left( \fft{R}{r}\right)^n P_n (\cos\theta)\,,
\ee
where $J_n$ are the multipole mass moments, $R$ is the equatorial radius of the sun and $P_n$ are the Legendre polynomials. Note that the effective mass in the first term is renormalized by the solar radiation pressure, whereas the multipole mass moments are not affected. We will consider only the case for which the orbit of the SSP satellite is confined to the equatorial plane. Then the acceleration is purely in the radial direction, given by
\be
a^r=-\fft{G\td M}{r^2}\left( 1-\fft{3MJ_2 R^2}{2\td Mr^2}+\fft{15MJ_4R^4}{8\td Mr^4}+\cdots\right)\,,
\ee
where we consider the quadrupole and octupole terms. Upon taking a power series expansion in the mass multipole moments, this leads to the relation
\be\label{oblateT}
T^2=\fft{4\pi^2r^3}{G\td M} \left[ 1+\fft{3MJ_2R^2}{2\td Mr^2}-\fft{3MR^4}{8\td Mr^4}\left( 5J_4+\fft{6MJ_2^2}{\td M}\right)+\cdots\right]\,.
\ee
From a model of the interior structure and of the solar rotation obtained from helioseismic inversions, the first few gravitational multipole moments have been theoretically predicted to be
\be
J_2\approx 2.2\times 10^{-7}\,,\qquad J_4\approx -4.5\times 10^{-9}\,,\qquad J_6\approx -2.8\times 10^{-10}\,,
\ee
which are in reasonable agreement with previous models \cite{oblateness2}. However, the quadrupole mass moment of the sun was recently been measured with unprecedented precision to be as much as $J_2\approx 9\times 10^{-6}$ during active phases of the solar cycle \cite{oblatesun}. 

First we consider the effects of the sub-leading term in (\ref{oblateT}), which is linear in $J_2$. Without the SRP, this latest value of $J_2$ increases the period by about $0.02$ s. With the SRP and the specifications given in (\ref{values}) for the SSP satellite, $J_2$ increases the period by about $105$ s. 

We now turn to the next highest terms in the power series expansion, which includes a linear term in the octupole mass moment $J_4$ and a quadratic term in the quadrupole mass moment $J_2$, as can be seen from (\ref{oblateT}). Without the SRP, the linear $J_4$ term dominates over the quadratic $J_2$ term and increases the period by $10^{-7}$ s. With the SRP, 
the linear $J_4$ term in (\ref{oblateT}) increases the period by $6\times 10^{-7}$ s. However, due to the SRP, the quadratic $J_2$ term now dominates over the linear $J_4$ term and leads to a decrease in the period by $0.003\ s$. Whatever the case, except for the linear $J_2$ term, the higher mass multipole terms have less of an effect on the period than even frame dragging.

\subsubsection{Non-circular orbits}

From the Newtonian equations for energy and angular momentum, and taking into account the mass quadrupole term in the gravitational potential given by (\ref{oblateV}), we find the orbital equation to be
\be
\left( \fft{dr}{d\phi}\right)^2=\left(\fft{E^2}{c^2}-c^2+\fft{2G\td M}{r}-\fft{L^2}{r^2}-\fft{GMJ_2R^2}{r^3}\right) \fft{r^4}{L^2}\,.
\ee
Note that this equation has the same form as the general relativistic orbital equation given by (\ref{curvedorbital}). Therefore, orbits which are not completely circular will undergo precession due to the oblateness of the sun. Using the coordinate $y$ given by (\ref{y}), we will consider a nearly circular orbit for which $y$ is small. Neglecting the $y^3$ terms, the orbital equation can be written as
\be
\left( \fft{dy}{d\phi}\right)^2=\fft{E^2/c^2+G^2\td M^2/L^2-c^2}{L^2}-\fft{G^4M_2\td M^3}{L^8}-\fft{3G^3M_2\td M^2}{L^6}y-\left( \fft{3G^2M_2\td M}{L^4}+1\right)y^2,
\ee
where $M_2\equiv MJ_2R^2$. The solution has the same form as (\ref{precsol}), where now the constants are given by
\bea
k &=& \sqrt{1+\fft{3G^2 M_2\td M}{L^4}}\,,\nn\\
y_0 &=& -\fft{3G^3 M_2\td M^2}{2L^6 k^2}\,,\nn\\
A &=& \fft{1}{k} \sqrt{\fft{E^2/c^2+G^2 \td M^2/L^2-c^2}{L^2}-\fft{GM_2\td M^3}{L^8}-\fft{3G^3 M_2\td M^2}{L^6}y_0-k^2y_0^2}\,.
\eea
The orbit oscillates about $y=y_0$ and the perihelion shift during one complete orbit is given by (\ref{pshift}). For small oblateness,
\be
\Delta\phi\approx -\ \fft{3\pi MJ_2R^2}{\td M r^2}\,,
\ee
where we have taken $L^2\approx G\td M r$. The characteristics of the precession due to the oblateness of the sun are rather different than the general relativistic precession. For instance, this precession occurs in the reverse direction relative to the orbit. For Mercury, we find the rate of precession to be about $-0.01$ arcseconds per year. Note that this is substantially less than the precession of Mercury's orbit due to the curvature of spacetime, which is about $+0.43$ arcseconds per year. For a conventional satellite in orbit at $r=0.05$ AU, we find the precession rate to be about $-14$ arcseconds per year. For our SPP satellite, the precession rate increases by a factor of $\sqrt{M/\td M}$ to about $-235$ arcseconds per year. It is rather interesting that the SRP increases the rate of precession due to the oblateness of the sun, whereas we found that the precession rate due to the curvature of spacetime is decreased by the SRP. Perhaps the SRP could be used to disentangle these two types of precession.

Since these two precessions occur in opposite directions, then for a certain choice of parameters they can actually cancel each other out. For a SSP satellite orbiting at $0.05$ AU, for example, this would occur if the solar sail has $\sigma\approx 0.00163$ kg/m$^2$. Of course, it is more feasible to vary the average orbital radius for a fixed $\sigma$ until the precessions cancel.

\subsection{Net electric charge on sun}

We will now consider the effect that a small amount of net charge $Q$ on the sun would have on an SSP satellite with charge $q$. It has been suggested that the sun has a net charge of  up to $Q\approx 77$ C \cite{solarcharge}. The spacetime is then described by the Reissner-Nordstr\"om metric, which has the form (\ref{form}), where the metric function $f$ is now given by
\be
f=1-\fft{2GM}{c^2r}+\fft{Gk_e Q^2}{c^4r^2}\,,
\ee
and the Coulomb constant $k_e=8.988\times 10^9$ N m$^2$/C$^2$.
Taking into account the electric force, we find that
\be
T^2\approx\fft{4\pi^2r^3}{G\td M} \left[ 1+\fft{\kappa}{c^2r}+\fft{k_eqQ}{Gm\td M}+\fft{k_eQ^2}{c^2\td M r}+\fft{k_e^2q^2 Q^2}{G^2m^2 \td M^2}\right]\,,
\ee
where we have kept only terms up to order $Q^2$. Note the $Q^2$ term in the period, due to the backreaction from the sun's charge on the geometry, is present even for a neutral SSP satellite. However, this term increases the period by an amount of only $10^{-35}$ s ($10^{-38}$ s without the SRP), which reflects the fact that the backreaction of the charge on the geometry is negligible, as to be expected. 

Primarily due to the photoelectric effect, but also the Compton effect and electron-positron pair production, a solar sail made from Beryllium, for example, will equilibrate to a charge per area 
of $0.065$ C/m$^2$ \cite{rk9,rk10}. For an SSP satellite with the specifications given in (\ref{values}) 
and a mass of $1000$ kg, this gives a charge $q=5\times 10^4$ C and an increase in 
period of about $230$ s ($0.05$ s without the SRP). Thus, due to the SRP, even 
a small charge $Q$ could certainly yield a measurable increase in the period, making this a 
potentially powerful test for net charge on the sun.

\subsection{Cosmological constant}

Since the SRP enhances a variety of small effects, including those associated with spacetime curvature, one could ask whether an SSP satellite could be used to test for the presence of a cosmological constant. In fact, supernovae observations suggest that our universe might have a very small positive cosmological constant  \cite{cosm1,cosm2}. In the presence of a cosmological constant $\Lambda$,
the spacetime in the vicinity of the sun is described by a metric of the form (\ref{form}), where the function $f$ is now given by
\be
f=1-\fft{2GM}{c^2r}-\fft{\Lambda r^2}{3}\,.
\ee
Considering the effect of $\Lambda$ for $r\gg GM/c^2$, we get
\be
T^2\approx\fft{4\pi^2r^3}{G\td M} \left[ 1+\fft{c^2\Lambda}{3G\td M}r^3\right]\,.
\ee
For $\Lambda\approx 10^{-52}$ m$^{-2}$ along with the specifications in (\ref{values}), this leads to an insignificant increase in period of $10^{-17}$ s ($10^{-21}$ s without the SRP). While one might be able specify values of $\sigma$ and $r$ such that the correction factor is actually significant, the period would then be too large to make this a feasible test for the presence of a cosmological constant.

\section{Circular orbits out of the plane of the sun}

\subsection{Review of non-Keplerian orbits}

\begin{figure}[ht]
   \epsfxsize=2.8in \centerline{\epsffile{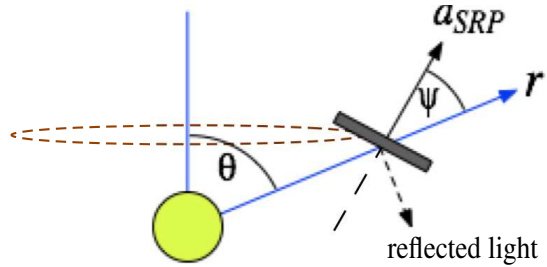}}
   \caption[FIG. \arabic{figure}.]{\footnotesize{Orienting a solar sail at an appropriate angle $\psi$ renders it possible to have a circular orbit outside of the plane of the sun.}}
\label{fig1}
\end{figure}

We will now briefly review non-Keplerian orbits within the framework of Newtonian gravity \cite{nonkepler1,nonkepler2,nonkepler3,rk1,nonkepler4}. The plane of a non-Keplerian orbit does not pass through the center of mass of the sun, and the SSP satellite is levitated above the sun, as shown in Figure \ref{fig1}. We work with spherical coordinates, where $r$ is the heliocentric distance and $\theta$ and $\phi$ are the polar and azimuthal angles, respectively. We orient our coordinate system such that the SSP orbits around a circle at constant $r$ and $\theta$. Then the orbital radius is given by $\rho=r \sin\theta$.

A non-Keplerian orbit can be maintained with a suitable pitch angle $\psi$ relative to the radially outgoing solar radiation hitting the surface of the sail. This enables one to control the direction of thrust due to the reflected portion of sunlight. By changing the pitch angle, one can also transfer the SSP satellite between different non-Keplerian orbits whose orbital planes have different orientations, as well as between non-Keplerian orbits and orbits within the plane of the sun. 

Note that while we are restricting ourselves to planar sails, for which the thrust is normal to the sail surface, the analysis for curved sails is similar as long as the sail curvature is symmetric about its center. 
The analysis below also carries over to non-Keplerian orbits of magnetic \cite{magneticsail1,magneticsail2} and electric \cite{electricsail1,electricsail2} sails, which rely on the solar wind rather than sunlight.

\subsubsection{Perfectly reflecting solar sail}

For a perfectly reflecting solar sail ($\eta=1$), the magnitude of the acceleration due to the solar radiation pressure is given by
\be
a_{SRP}=\fft{\kappa \cos\psi}{r^2}\,.
\ee
The factor of $\cos\psi$ is present because we have projected the SRP along the normal direction to the solar sail. We can break this acceleration into its components expressed in spherical coordinates as follows:
\be
a_{SRP}^r=a_{SRP} \cos\psi\,,\qquad a_{SRP}^{\theta}=-a_{SRP}\sin\psi\,,\qquad a_{SRP}^{\phi}=0\,,
\ee
where $r$ is the heliocentric distance, $\theta$ is the polar angle and $\phi$ is the azimuthal angle. Including the contribution of acceleration due to the gravitational attraction of the sun gives
\be\label{a1}
a^r=\fft{\kappa\cos^2\psi-GM}{r^2}\,,\qquad a^{\theta}=-\fft{\kappa}{r^2} \cos\psi\sin\psi\,,\qquad a^{\phi}=0\,.
\ee
Acceleration in terms of spherical coordinates can generally be written as
\bea\label{a2}
a^r &=& \ddot r-r\dot\theta^2-r \sin^2\theta\ \dot\phi^2\,,\nn\\
a^{\theta} &=& r\ddot\theta+2\dot r\dot\theta-r\sin\theta\cos\theta\ \dot\phi^2\,,\nn\\
a^{\phi} &=& r^{-1} \partial_t (r^2\sin^2\theta\dot\phi)\,,
\eea
where $\dot{}\equiv\partial_t$. Equating the components of acceleration in (\ref{a1}) and (\ref{a2}) gives the equations of motion for the solar sail. For constant $r$ and $\theta$, these equations reduce to
\bea\label{eom}
a^r &=& -r\sin^2\theta\ \dot\phi^2=-\fft{G\td M}{r^2}\,,\nn\\
a^{\theta} &=& -r\sin\theta\cos\theta\ \dot\phi^2=-\fft{\kappa}{r^2} \cos\psi\sin\psi\,,\nn\\
a^{\phi} &=& r^{-1} \partial_t (r^2\sin^2\theta\ \dot\phi)=0\,,
\eea
where
\be
\td M\equiv M-\fft{\kappa}{G}\cos^2\psi\,.
\ee
The $\phi$ equation implies that $r^2\sin^2\theta\ \dot\phi=L_N$ is a constant of motion, which corresponds to the angular momentum per unit mass of the solar sail. As before, we will use the subscript $N$ to denote that this is a conserved quantity within the Newtonian approximation.

\begin{figure}[ht]
\begin{center}
$\begin{array}{c@{\hspace{0.5in}}c}
\epsfxsize=2.6in \epsffile{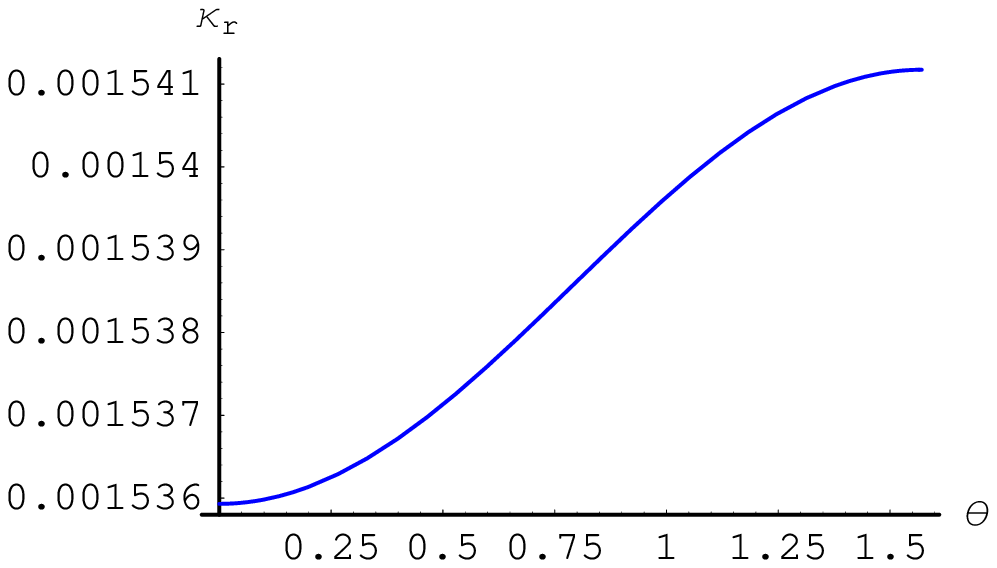} &
\epsfxsize=2.6in \epsffile{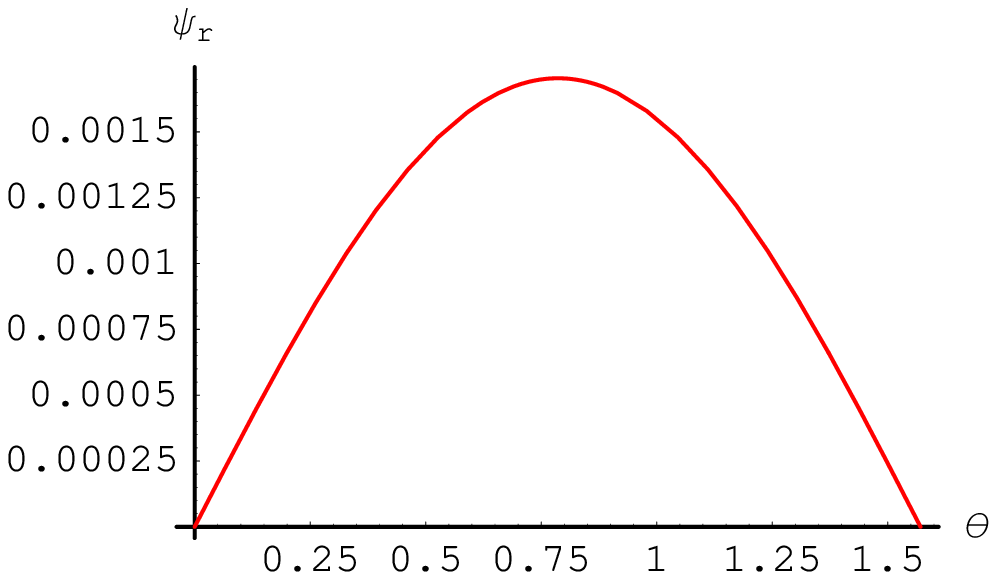}
\end{array}$
\end{center}
\caption[FIG. \arabic{figure}.]{\footnotesize{$\kappa_r$ and $\psi_r$ as functions of $\theta$ for the SSP satellite.}}
   \label{fig4}
\end{figure}

We can write $\dot\phi=2\pi/T$, where $T$ is the period of a complete orbit. Substituting this into the $a^r$ equation in (\ref{eom}) enables us to express the period in terms of the heliocentric distance and polar angle as
\be\label{T1}
T^2=\fft{4\pi^2}{G\td M}\ r^3\sin^2\theta\,,
\ee
which bears some similarity to Kepler's third law, in that $T\sim \rho^{3/2}$. From (\ref{T1}) and the $a^{\theta}$ equation in (\ref{eom}), we find that 
\bea\label{kpsin}
\kappa_r &=& GM-\fft{4\pi^2r^3}{T^2}+\fft{4\pi^2GMr^3\cos^2\theta}{GMT^2-4\pi^2 r^3\sin^2\theta}\,,\nn\\
\tan\psi_r &=& \fft{2\pi^2 r^3 \sin 2\theta}{GMT^2-4\pi^2 r^3 \sin^2\theta}\,,
\eea
where the subscript $r$ indicates that these are the parameters for a perfectly reflecting solar sail. Thus, the pitch angle $\psi_r$ and the $\kappa_r$ parameter are both determined by the period, heliocentric distance and polar angle of the orbit. For fixed $r=0.5$ AU and $T=70$ days, $\kappa_r$ and $\psi_r$ are shown in Figure \ref{fig4} for $0\le\theta\le\pi/2$. Note that for $\pi/2\le\theta\le\pi$, the plot of $\kappa_r$ is flipped horizontally and that of $\psi_r$ is flipped vertically. 
$\kappa_r$ monotonically increases with $\theta$ and reduces to the $\kappa$ given by (\ref{k1}) for $\theta=\pi/2$. $\psi_r$ is at a maximum for $\theta=\pi/4$ and vanishes for $\theta=0$ and $\pi/2$. At first sight, it might seem odd that $\psi_r$ decreases with $\theta$ in the region $\pi/4\le\theta\le\pi/2$. However, from Figure \ref{fig1} we see that the angle of the normal direction of the solar sail relative to the line passing through the poles of the sun is given by $\theta-\psi_r$, which monotonically increases with $\theta$ down to the equatorial plane, as expected.

\subsubsection{Partially absorbing solar sail}

We will now consider the more realistic scenario in which a portion of the light is absorbed by the surface of the solar sail, which has also been considered quite a while ago in \cite{nonkepler1,nonkepler2}. We saw above that reflected light pushes the solar sail at an angle $\psi$ relative to the radial direction. For a partially absorbing solar sail, the absorbed light pushes the solar sail radially outwards. The fraction of light reflected is $2\eta-1$ and the fraction of light absorbed is $2(1-\eta)$. Putting all of the contributions to the acceleration together, we find that
\be\label{a3}
a^r=\fft{(1-\eta)\td\kappa+(2\eta-1)\td\kappa\cos^2\psi-GM}{r^2}\,,\quad a^{\theta}=-\fft{(2\eta-1)\td\kappa}{r^2} \cos\psi\sin\psi\,,\quad a^{\phi}=0\,,
\ee
where $\td\kappa\equiv L_s/(2\pi c\sigma)$. This reduces to (\ref{a1}) for $\eta=1$. Equating the components of acceleration in (\ref{a2}) and (\ref{a3}) and taking $r$ and $\theta$ to be constant gives
\bea\label{eom3}
a^r &=& -r\sin^2\theta\ \dot\phi^2=-\fft{G\td M^{\ast}}{r^2}\,,\nn\\
a^{\theta} &=& -r\sin\theta\cos\theta\ \dot\phi^2=-\fft{(2\eta-1)\td\kappa}{r^2} \cos\psi\sin\psi\,,\nn\\
a^{\phi} &=& r^{-1} \partial_t (r^2\sin^2\theta\ \dot\phi)=0\,,
\eea
where we now have the effective solar mass as
\be\label{MM}
\td M^{\ast}\equiv M-\fft{\td\kappa}{G}\left[ (1-\eta)+(2\eta -1)\cos^2\psi\right]\,.
\ee
Plugging $\dot\phi=2\pi/T$ into the $a^r$ equation in (\ref{eom3}) yields the expression (\ref{T1}) with $\td M$ replaced by $\td M^{\ast}$. Combining this with the $a^{\theta}$ equation in (\ref{eom3}) gives
\bea\label{kpsia}
\tan\psi_a &=& \fft{(2\eta-1)(GMT^2-4\pi^2r^3\sin^2\theta)}{4\pi^2(1-\eta)r^3\sin(2\theta)} \left[ 1-\sqrt{1+\fft{\eta(\eta-1)(4\pi^2r^3\sin2\theta)^2}{(2\eta-1)^2 (GMT^2-4\pi^2r^3\sin^2\theta)^2}}\right]\,,\nn\\
\td\kappa_a &=& \fft{(GMT^2-4\pi^2 r^3\sin^2\theta)(1+\tan^2\psi_a)}{T^2\left[\eta+(1-\eta)\tan^2\psi_a\right]}\,,
\eea
where the subscripts $a$ indicate that a portion of the light is absorbed by the solar sails. Consider the case in which the solar sail is close to being a perfect reflector. Then $\ep\equiv 1-\eta\ll 1$ and we have
\be
(\tan\psi_a, \td\kappa_a)=(\tan\psi_r, \kappa_r)\times  \left[ 1+\left( 1+\fft{(2\pi^2 r^3 \sin 2\theta)^2}{(GMT^2-4\pi^2r^3\sin^2\theta)^2}\right) \ep\right].
\ee
$\tan\psi_r$ and $\kappa_r$ are given by (\ref{kpsin}). As one might expect, the nonzero absorption of light means that we must have a larger pitch angle $\psi$ in order to have a non-Keplerian orbit. 

\begin{figure}[ht]
\begin{center}
$\begin{array}{c@{\hspace{0.5in}}c}
\epsfxsize=2.6in \epsffile{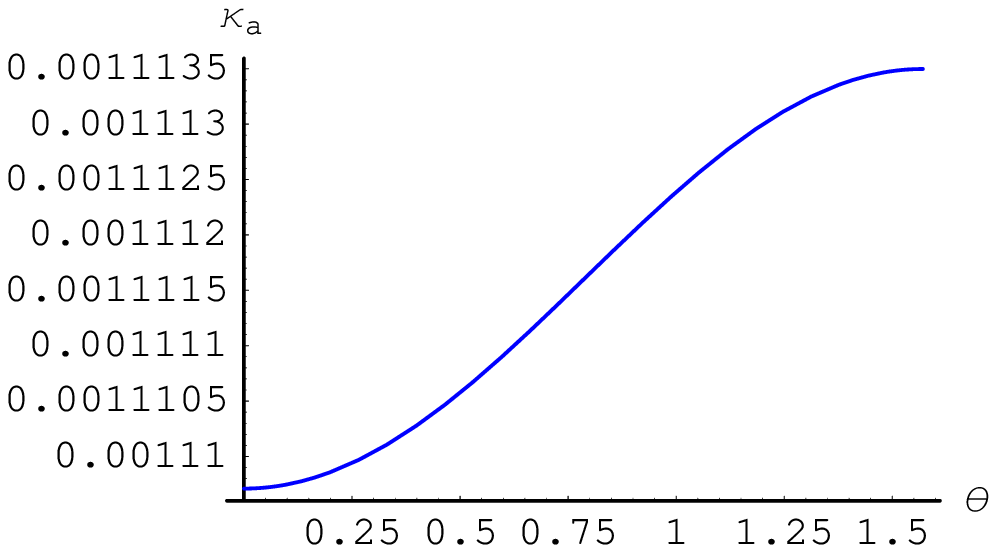} &
\epsfxsize=2.6in \epsffile{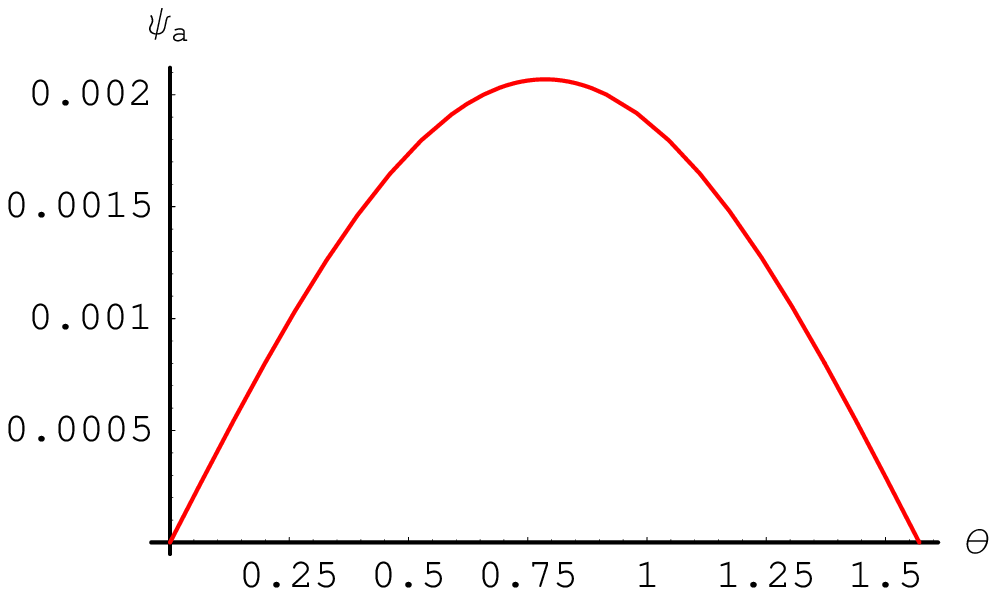}
\end{array}$
\end{center}
\caption[FIG. \arabic{figure}.]{\footnotesize{$\kappa_a$ and $\psi_a$ as functions of $\theta$ for the SSP satellite.}}
   \label{fig5}
\end{figure}

For fixed $r=0.5$ AU, $T=70$ days and $\eta=0.85$, $\kappa_a$ and $\psi_a$ are shown in Figure \ref{fig5} for $0\le\theta\le\pi/2$.
The partial absorption of light by the surface of the solar sail causes $\kappa_a$ to decrease and $\psi_a$ to increase, which can be seen for $\eta=0.85$ by comparing Figures \ref{fig4} and \ref{fig5}. We have confirmed that these changes in $\kappa_a$ and $\psi_a$ occur monotonically in $\eta$ for $0.5<\eta\le 1$. The $\theta$ dependence of $\kappa_a$ and $\psi_a$ nevertheless remain qualitatively the same, regardless of whether any light is absorbed by the sail or not. This agrees with the conclusions of \cite{nonkepler1,nonkepler2}

\subsection{Static curved spacetime}

We will now consider the effects of curved spacetime on non-Keplerian orbits. We will begin in this section by considering the sun to be a static central mass, and ignore the frame dragging due to its rotation. Then the exterior spacetime of the sun is approximately described by the Schwarzschild metric, which is given by (\ref{form}) and (\ref{f}). 

The acceleration 4-vector of an object moving within this background can be obtained by taking the covariant derivative of the velocity 4-vector $u^{\mu}=dx^{\mu}/d\tau$ as follows:
\be\label{covarianta}
a^{\mu}=\fft{du^{\mu}}{d\tau}+\Gamma_{\alpha\beta}^{\mu} u^{\alpha}u^{\beta}\,,\qquad \Gamma_{\alpha\beta}^{\mu}=\ft12g^{\mu\nu}(g_{\nu\alpha,\beta}+g_{\nu\beta,\alpha}-g_{\alpha\beta,\nu})\,,
\ee
where $x^{\mu}=(t,r,\theta,\phi)$ and $\tau$ is the proper time. We find that
\bea\label{gra1}
a^t &=& f^{-1} \partial_{\tau} (f\dot t)\,,\nn\\
a^r &=& \ddot r+\fft{GM}{r^2}\ f\dot t^2-\fft{GM}{c^2r^2}\ f^{-1}\dot r^2-fr(\dot\theta^2+\sin^2\theta\ \dot\phi^2)\,,\nn\\
a^{\theta} &=& r\ddot\theta+2\dot\theta\dot r-r\sin\theta\cos\theta\ \dot\phi^2\,,\nn\\
a^{\phi} &=& r^{-1} \partial_{\tau} (r^2\sin^2\theta\ \dot\phi)\,.
\eea
Note that the spatial components of the acceleration in (\ref{gra1}) reduce to those in (\ref{a2}) in the Newtonian limit $c^2r>>GM$.

The acceleration due to the solar radiation pressure is given by
\be\label{gra2}
a^r=\fft{(1-\eta)\td\kappa+(2\eta-1)\td\kappa\cos^2\psi}{r^2}\,,\qquad a^{\theta}=-\fft{(2\eta-1)\td\kappa}{r^2} \cos\psi\sin\psi\,,\qquad a^t=a^{\phi}=0\,.
\ee
Equating the components of the acceleration in (\ref{gra1}) and (\ref{gra2}) gives the equations of motion for the solar sail. For constant $\theta$, these equations reduce to
\bea\label{eom2}
a^t &=& f^{-1} \partial_{\tau} (f\dot t)=0\,,\nn\\
a^r &=& \ddot r+\fft{GM}{r^2}\ f\dot t^2-\fft{GM}{c^2r^2}\ f^{-1}\dot r^2-fr\sin^2\theta\ \dot\phi^2=\fft{(1-\eta)\td\kappa+(2\eta-1)\td\kappa\cos^2\psi}{r^2}\,,\nn\\
a^{\theta} &=& -r\sin\theta\cos\theta\ \dot\phi^2=-\fft{(2\eta-1)\td\kappa}{r^2} \cos\psi\sin\psi\,,\nn\\
a^{\phi} &=& r^{-1} \partial_{\tau} (r^2\sin^2\theta\ \dot\phi)=0\,.
\eea
The $t$ and $\phi$ equations imply that 
\be\label{constants}
f\dot t=E/c^2\,,\qquad r^2\sin^2\theta\ \dot\phi=L\,,
\ee
are constants of motion. The first integral of the $r$ equation gives
\be
\dot r^2=\fft{E^2}{c^2}-c^2+\fft{2G\td M^{\ast}}{r}-\fft{L^2}{r^2}f\,,
\ee
where $\td M^{\ast}$ is defined in (\ref{MM}).

Circular orbits must satisfy the conditions given in (\ref{circular}), which yields
\be
E^2 = c^4+\left( \fft{4GM-c^2r}{c^2r-3GM}\right) \fft{c^2G\td M^{\ast}}{r},\quad
L^2 = \fft{c^2G\td M^{\ast} r^2}{c^2r-3GM}.
\ee
Using (\ref{constants}), we find the orbital period $T$ to be given the expression
\be
T^2=\fft{4\pi^2}{G\td M^{\ast}}\sin^2\theta\ r^3 \left[ 1+\td\kappa \left[(1-\eta)+(2\eta-1)\cos^2\psi\right] \fft{(c^2r-4GM)}{(c^2r-2GM)^2}\right]\,.
\ee
Keeping only the leading general relativistic correction gives
\be\label{Tgr}
T^2\approx\fft{4\pi^2}{G\td M^{\ast}}\sin^2\theta\ r^3 \Big[ 1+\fft{\td\kappa \left[(1-\eta)+(2\eta-1)\cos^2\psi\right]}{c^2r}\Big]\,.
\ee
From (\ref{Tgr}) and the $a^{\theta}$ equation in (\ref{eom2}), we find the following equation for the pitch angle $\psi$:
\bea
&& \fft{(2\eta-1)c^2\tan\psi (GMT^2-4\pi^2 r^3\sin^2\theta)}{2\pi^2r^3\sin 2\theta [(1-\eta)(1+\tan^2\psi)+2\eta-1]}-c^2= \nn\\ && \fft{3GM}{r}+\fft{4\pi^2r^2\sin^2\theta}{T^2}+\fft{2\pi^2r^2\sin 2\theta [(1-\eta)(1+\tan^2\psi)+2\eta-1]}{(2\eta-1)T^2\tan\psi}\,.
\eea
Since all of the general relativistic corrections on the right-hand side go as $c^{-2}$, we can solve for $\psi$ perturbatively as $\psi=\psi_a+c^{-2}\psi_1$. Then we find
\bea\label{grpsikappa}
\tan\psi &=& \tan\psi_a+\fft{1}{c^2} (1+\tan^2\psi_a)\tan\psi_1,\nn\\
\td\kappa &=& \td\kappa_a \left( 1+\fft{2(2\eta-1)\tan\psi_a\tan\psi_1}{c^2[\eta+(1-\eta)\tan^2\psi_a]}-\fft{4\pi^2r^2\sin^2\theta}{c^2T^2}\right).
\eea
where $\tan\psi_a$ and $\td\kappa_a$ are given in (\ref{kpsia}) and
\bea
\tan\psi_1 &=& \fft{2\pi^2r^3\sin 2\theta [\eta+(1-\eta)\tan^2\psi_a]}{(2\eta-1)(1+\tan^2\psi_a)(GMT^2-4\pi^2r^3\sin^2\theta)[\eta-(1-\eta)\tan^2\psi_a]}\nn\\
&& \times \left( \fft{GM}{r}+\fft{4\pi^2r^2\sin^2\theta}{T^2}+\fft{2\pi^2r^2\sin 2\theta [\eta+(1-\eta)\tan^2\psi_a]}{(2\eta-1)T^2\tan\psi_a}\right).
\eea
For a perfectly reflecting solar sail, (\ref{grpsikappa}) reduces to
\bea
\tan\psi &=& \tan\psi_r+\fft{4\pi^2GMr^2\sin 2\theta}{c^2 (GMT^2-4\pi^2 r^3 \sin^2\theta)}\,,\nn\\
\td\kappa &=& \td\kappa_r \left( 1+\fft{GM(4\pi^2r^3\sin 2\theta)^2}{c^2r[(GMT^2)^2+8\pi^2r^3\sin^2\theta (2\pi^2r^3-GMT^2)]}-\fft{4\pi^2 r^2\sin^2\theta}{c^2T^2}\right)\,,
\eea
where $\tan\psi_r$ and $\td\kappa_r$ are given in (\ref{kpsin}). We find that the curvature of spacetime has an extremely small effect on $\sigma$ and $\psi$. For instance, for $r=0.5$ AU, $T=70$ days and $\theta=45^{\circ}$, $\sigma$ increases by about $4\times 10^{-13}$ kg/m$^2$ and $\psi$ increases by about $4\times 10^{-8}$ degrees.

\subsection{Frame dragging}

\subsubsection{Orbits parallel to the equatorial plane of the sun}

As is the case for orbits within the plane of the sun, non-Keplerian orbits will also be affected by frame dragging due to the sun's rotation. We will first consider the effect of frame dragging on non-Keplerian orbits which are parallel to the equatorial plane of the sun.

The exterior spacetime of the sun is approximately described by the large-distance limit of the Kerr metric, which is given by (\ref{kerrmetric}). From (\ref{covarianta}), we find that the components of the acceleration 4-vector of an object moving within this background are
\bea\label{kerra1}
a^t &=& f^{-1} \partial_{\tau} (f\dot t)+\fft{2GJ}{c^4f} \dot\phi\ \partial_{\tau} \left( \fft{\sin^2\theta}{r}\right)
\,,\nn\\
a^r &=& \ddot r+\fft{GM}{r^2}\ f\dot t^2-\fft{GM}{c^2r^2}\ f^{-1}\dot r^2-fr(\dot\theta^2+\sin^2\theta\ \dot\phi^2)-\fft{2GJ}{c^2r^2} f\dot\phi\dot t \sin^2\theta\,,\nn\\
a^{\theta} &=& r\ddot\theta+2\dot\theta\dot r-r\sin\theta\cos\theta\ \dot\phi^2+\fft{4GJ}{c^2r^2} \dot\phi\dot t \sin\theta \cos\theta\,,\nn\\
a^{\phi} &=& r^{-1} \partial_{\tau} (r^2\sin^2\theta\ \dot\phi)+\fft{2GJ}{c^2r^2}\dot t\left( \fft{\sin\theta}{r}\dot r-2\dot\theta\cos\theta\right)\,.
\eea
Note that the $a^{\mu}$ reduce to what is given by (\ref{gra1}) in the limit $J=0$. Equating the components of the acceleration in (\ref{kerra1}) and (\ref{gra2}) gives the equations of motion for the solar sail. For constant $\theta$, the equations for $a^r$ and $a^{\theta}$ reduce to
\bea\label{eom4}
a^r &=& \ddot r+\fft{GM}{r^2}\ f\dot t^2-\fft{GM}{c^2r^2}\ f^{-1}\dot r^2-fr\sin^2\theta\ \dot\phi^2-\fft{2GJ}{c^2r^2} f\dot\phi\dot t \sin^2\theta\nn\\
&=& \fft{(1-\eta)\td\kappa+(2\eta-1)\td\kappa\cos^2\psi}{r^2}\,,\nn\\
a^{\theta} &=& -r\sin\theta\cos\theta\ \dot\phi^2+\fft{4GJ}{c^2r^2} \dot\phi\dot t \sin\theta \cos\theta=-\fft{(2\eta-1)\td\kappa}{r^2} \cos\psi\sin\psi\,.
\eea
The first integral of the $r$ equation gives
\be
\dot r^2=\fft{E^2}{c^2}-c^2+\fft{2G\td M^{\ast}}{r}-\fft{L^2}{r^2}f-\fft{4GJEL\sin\theta}{c^4r^3}\,,
\ee
where $\td M^{\ast}$ is defined in (\ref{MM}).

Circular orbits must satisfy the conditions given in (\ref{circular}), which yields
\bea\label{kerrEL}
E^2 &=& c^4+\left( \fft{4GM-c^2r}{c^2r-3GM}\right) \fft{c^2G\td M^{\ast}}{r}-\fft{2c^2GJ\sin\theta}{r(c^2r-3GM)^2} \sqrt{\fft{G\td M^{\ast} X}{r}}\,,\nn\\
L^2 &=&  \fft{c^2G\td M^{\ast} r^2}{c^2r-3GM}-\fft{6GJ\sin\theta}{(c^2r-3GM)^2} \sqrt{G\td M^{\ast} r X}\,.
\eea
where
\be
X \equiv c^4r^2-c^2G(3M+\td M^{\ast})r+4G^2M\td M^{\ast}\,.
\ee
Using (\ref{pt}) and (\ref{kerrEL}), we find the orbital period $T$ to be given by
\bea
T^2 &=& \fft{4\pi^2}{G\td M^{\ast}}\sin^2\theta\ r^3 \left[ 1+\td\kappa \left[(1-\eta)+(2\eta-1)\cos^2\psi\right] \fft{(c^2r-4GM)}{(c^2r-2GM)^2}\right]\nn\\
&\times& \left[ 1+\fft{2GJ\sin\theta}{c^2r}\left( \fft{L}{r^2E}+\fft{E}{c^2fL}\right)\right]\,.
\eea
As we did for the case of orbits within the plane of the sun, we have presented $T^2$ as a product of three factors. The first factor is the same as for non-Keplerian orbits in the Newtonian approximation for gravity. The second factor is due to the simultaneous effects of the SRP and the curvature of spacetime outside of a static central body. The third factor embodies the combined effects of the SRP and frame dragging due to the rotation of the sun. 

Keeping only the leading correction due to spacetime curvature gives
\be\label{nonkeplerTkerr}
T^2\approx\fft{4\pi^2}{G\td M^{\ast}}\sin^2\theta\ r^3 \Big[ 1+\fft{\td\kappa \left[(1-\eta)+(2\eta-1)\cos^2\psi\right]}{c^2r}\Big]\left[ 1+\fft{2\sqrt{G}J\sin\theta}{c^2\sqrt{\td M^{\ast}} r^{3/2}}\right]\,.
\ee
As was the case for orbits within the plane of the sun, the SRP increases the effect that frame dragging has on the orbital period, since it is the renormalized mass $\td M$ that appears in the last factor. 
Although the solar sail parameter $\td\kappa$ and the pitch angle $\psi$ could be determined in terms of the other parameters from (\ref{nonkeplerTkerr}) and the $a^{\theta}$ equation in (\ref{eom4}), frame dragging has a negligible effect on $\td\kappa$ and $\psi$.

\subsubsection{Polar orbits outside of the plane of the sun}

\begin{figure}[ht]
   \epsfxsize=2.5in \centerline{\epsffile{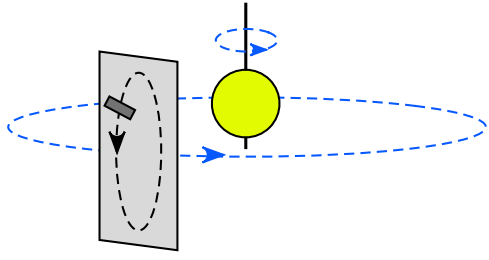}}
   \caption[FIG. \arabic{figure}.]{\footnotesize{The Lense-Thirring effect for non-Keplerian polar orbits.}}
\label{fig2}
\end{figure}

We will now consider the effect of frame dragging on non-Keplerian orbits which are parallel to polar orbits, and outside of the plane of the sun. The plane of non-Keplerian polar orbits undergoes the Lense-Thirring effect, as shown in Figure \ref{fig2}.

We do not present the resulting equations since they are rather cumbersome. However, recall from the discussion of frame dragging for polar orbits within the plane of the sun that, up to linear order in $J$, the precession frequency was not affected by the SRP nor the polar orbital motion of the satellite, and depended only on the heliocentric distance $r$ and not the polar angle $\theta$. Up to leading order in $J$, these characteristics also hold for non-Keplerian orbits and the precession frequency is given approximately by (\ref{precession}). Thus, in one orbital period given approximately by (\ref{Tgr}), the angle of precession is
\be
\Delta\phi\approx \fft{4\pi GJ\sin\theta}{c^2\sqrt{G\td M^{\ast}r^3}}\,.
\ee

\subsection{Oblateness of the sun}

We will now consider the effect of the quadrupole mass moment of the sun on non-Keplerian orbits. The gravitational potential is given by (\ref{oblateV}), from which we find that the components of acceleration due to the gravitational potential are
\be
a^r = -\fft{GM}{r^2}-\fft{3GMJ_2R^2}{2r^4}\ (3\cos^2\theta -1)\,,\quad
a^{\theta} = -\fft{3GMJ_2R^2}{2r^4}\sin (2\theta)\,,\quad
a^{\phi} = 0\,.
\ee
Including the contribution of acceleration due to the SRP gives
\bea
a^r &=& -\fft{G\td M^{\ast}}{r^2}-\fft{3GMJ_2R^2}{2r^4}\ (3\cos^2\theta -1)\,,\nn\\
a^{\theta} &=& -\fft{(2\eta-1)\td\kappa}{2r^2}\sin (2\psi)-\fft{3GMJ_2R^2}{2r^4}\sin (2\theta)\,,\nn\\ 
a^{\phi} &=& 0\,,
\eea
where $\td\kappa\equiv L_s/(2\pi c\sigma)$ and $\td M^{\ast}$ is given by (\ref{MM}). Equating the above components of acceleration to those in (\ref{a2}) and taking $r$ and $\theta$ to be constant gives
\bea\label{oblateartheta}
a^r &=& -r\sin^2\theta\ \dot\phi^2= -\fft{G\td M^{\ast}}{r^2}-\fft{3GMJ_2R^2}{2r^4}\ (3\cos^2\theta -1)\,,\nn\\
a^{\theta} &=& -r\sin\theta\cos\theta\ \dot\phi^2=-\fft{(2\eta-1)\td\kappa}{2r^2}\sin (2\psi)-\fft{3GMJ_2R^2}{2r^4}\sin (2\theta)\,,\nn\\
a^{\phi} &=& r^{-1} \partial_t (r^2\sin^2\theta\ \dot\phi)=0\,,
\eea
Substituting $\dot\phi=2\pi/T$ into the $a^r$ equation yields
\be
T^2\approx \fft{4\pi^2r^3\sin^2\theta}{G\td M^{\ast}} \left( 1+\fft{3MJ_2R^2}{2\td M^{\ast}r^2}(1-3\cos^2\theta )\right)\,.
\ee
Note that this reduces to the corresponding terms in (\ref{oblateT}) when $\theta=\pi/2$. Interestingly enough, the oblateness of the sun increases the period for orbits within the angular range of approximately $55^{\circ}<\theta <90^{\circ}$, whereas the period is decreased for $\theta < 55^{\circ}$.

From the $a^r$ and $a^{\theta}$ equations in (\ref{oblateartheta}), we find
\bea
\kappa [1-\eta+(2\eta-1)\cos^2\psi] &=& GM-\fft{4\pi^2 r^3\sin^2\theta}{T^2}+\fft{3GMJ_2R^2}{2r^2} (3\cos^2\theta-1)\,,\nn\\
(2\eta-1)\kappa \sin (2\psi) &=& \left( \fft{4\pi^2r^3}{T^2}-\fft{3GMJ_2R^2}{r^2}\right) \sin (2\theta)\,,
\eea
from which one can express $\kappa$ and the pitch angle $\psi$ in terms of the orbital parameters and $\eta$. Since these expressions are rather long, we present the idealized case in which the solar sail is a perfect reflector:
\bea
\kappa &\approx& \kappa_r-\fft{3GMJ_2R^2}{2r^2}\left[ 1-3\cos^2\theta+4\pi^2 r^3 \sin^2 (2\theta) \left( \fft{GMT^2+\pi^2 r^3 (7\cos^2\theta -5)}{(GMT^2-4\pi^2 r^3 \sin^2\theta)^2}\right) \right]\,,\nn\\
\tan\psi &\approx& \tan\psi_r \left[ 1-\fft{3GMJ_2R^2T^2}{4\pi^2 r^5} \left( \fft{GMT^2+2\pi^2r^3(5\cos^2\theta-3)}{GMT^2-4\pi^2r^3\sin^2\theta}\right)\right]\,,
\eea
where we have included the leading term in $J_2\ll 1$. We find that the oblateness of the sun has an extremely small effect on $\sigma$ and $\psi$. For instance, for $r=0.5$ AU, $T=70$ days and $\theta=45^{\circ}$, $\sigma$ decreases by about $2\times 10^{-12}$ kg/m$^2$ and $\psi$ decreases by about $2\times 10^{-7}$ degrees.

\section{Conclusions}

We have considered the effects of various phenomena along with the solar radiation pressure (SRP) on bound orbits of a solar sail propelled (SSP) satellite. The consideration of the SRP on its own leads to a renormalization of the effective solar mass. However, when the SRP is coupled with other effects, such as spacetime curvature, then the resulting orbital characteristics depend on the solar mass $M$ and the solar sail parameter $\kappa$ separately, and not simply in the combination of a renormalized mass.

For simplicity, we first considered the effects on the period of a circular orbit in the plane of the sun. In the table below, we summarize our results for the change in period $\Delta T$ of an SSP satellite with the specifications given in (\ref{values}).
\vspace{.4cm}
\begin{center}
\begin{tabular}{|c|c|c|}
\hline Phenomenon & $\Delta T$ without SRP & $\Delta T$ with SRP \\
\hline\hline Static Curvature of Spacetime & $0$ s & $0.6$ s \\
\hline Frame Dragging & $4\times 10^{-5}$ s & $0.01$ s \\
\hline Quadrupole Mass Moment of Sun & $0.02$ s & $105$ s \\
\hline Octupole Mass Moment of Sun & $10^{-7}$ s & $6\times 10^{-4}$ s \\
\hline Net Charge of Sun & $0.05$ s & $230$ s \\
\hline Cosmological Const. & $10^{-21}$ s & $10^{-17}$ s \\
\hline
\end{tabular}
\end{center}
\vspace{.4cm}
Recall that the base period without the SRP is about 4 days, and with the SRP is about 70 days. The SRP generally augments the change in period due to various phenomena by a factor of 
about $1000$ or more. This might be sufficient to observe the $\Delta T$ from some phenomena, such as the quadrupole mass moment of the sun and a possible net charge on the sun. 

Although we have isolated the effects of each of these phenomena, the reality of course is that they all have simultaneous effects on the period. For instance, to compute the combined effects of frame dragging and the mass multipole moments of the sun on orbits, one could use the metric which describes a rotating deformed mass with all mass multipole moments \cite{together}. Naturally, the challenge is to isolate the various effects from one another, especially since the SRP would also enhance additional effects that we have not discussed, such as the gravitational effects of the planets. 

As an example, consider the effect on the period due to the static curvature of spacetime and the SRP. Note that this has no counterpart without the SRP. Namely, when the SRP is not taken into account, the static curvature of spacetime does not alter the period as given by Newtonian gravity. Unfortunately, this rather interesting effect could likely be overshadowed by the dominant effect of the quadrupole mass moment of the sun. However, the effect of the mass moments of the sun depends on the plane of orbit, whereas the effect of the static curvature of spacetime does not. Thus, considering different orbital planes could aid in isolating the weaker effect of spacetime curvature. We have also confirmed that the higher mass moments of the sun have negligible effect compared to that of the static curvature of spacetime. 

Likewise, for a charged satellite, the effects of net charge on the sun could be smeared out by the sun's magnetic field. This too is unfortunate, since the magnitude of the net charge on the sun is not known very accurately, so any improved measurements of this are highly desirable. Since the magnetic force depends on the velocity of the satellite, the situation might be substantially improved by considering orbits with various orientations and speeds. However, since the magnetic field of the sun is highly variable, extracting the effects of net electric charge remains a universal problem.

We also considered orbits which deviated from being perfectly circular. There are two sources of perihelion shift. The curvature of spacetime causes the perihelion to be shifted in the direction of the orbit, whereas the oblateness of the sun causes the perihelion to be shifted in the reverse direction. Without the effects of the SRP, the perihelion shift due to spacetime curvature is generally larger than that due to the oblateness of the sun. However, this is not the case when the effects of the SRP are taken into account on an SSP satellite. As shown in the table below, the SRP decreases the perihelion shift due to spacetime curvature but {\it increases} the perihelion shift due to the oblateness of the sun. 
\vspace{.4cm}
\begin{center}
\begin{tabular}{|c|c|c|}
\hline Cause of Perihelion Shift & $\Delta\phi$ without SRP & $\Delta\phi$ with SRP \\
\hline\hline Spacetime Curvature & $70^{\prime\prime}/$year & $4^{\prime\prime}/$year \\
\hline Oblateness of Sun & $-14^{\prime\prime}/$year & $-235^{\prime\prime}/$year \\
\hline
\end{tabular}
\end{center}
\vspace{.4cm}
Thus, SSP satellites might enable one to disentangle the perihelion shift due to the oblateness of the sun from the general relativistic perihelion shift by making the former one more pronounced. Of course, these are both small effects which would need to be isolated from a plethora of other types of effects.

We have also investigated circular orbits outside of the plane of the sun, which are commonly referred to as non-Keplerian orbits. Such orbits are possible for a solar sail with an appropriate design parameter $\kappa$ and pitch angle $\psi$, both of which depend on the specified orbital period, radius and distance from the plane of the sun. We have computed how $\kappa$ and $\psi$ depend on the heliocentric distance and polar angle of the orbit, as well as on partial absorption of light by the surface of the solar sail. We have also considered how $\kappa$ and $\psi$ are altered due to general relativistic effects as well as the oblateness of the sun. 

For both orbits within the plane of the sun as well as non-Keplerian orbits, we have considered the Lense-Thirring effect, which is the precession of polar orbits. To leading order, the rate of precession is not altered by the SRP. Namely, any satellite in polar orbit at a distance from the sun of $0.05$ AU will experience a rate of precession of about $0.03$ arcseconds per year.

For the case of perfectly reflecting solar sails, types of non-Keplerian orbits have been classified and their linear stability characteristics have been studied \cite{rk1}. In particular, stable and unstable regions in space have been plotted. It would be interesting to see how the boundary between stability and instability is altered due to various effects such as the partial absorption of light, the oblateness of the sun and curved spacetime. The patching between two non-Keplerian orbits of different orientations, as well as between a non-Keplerian orbit and a Keplerian orbit, has also been studied \cite{rk1}. It would be interesting to see if the allowed types of patching are changed by these various effects.

In this study, we have assumed that the intensity of solar radiation goes as the inverse square of the heliocentric distance. However, this assumption is not valid when the finite angular size of the sun is taken into account-- along with the oblateness of the sun, sunspots, etc. Although the deviation from an inverse square law is certainly small, orbits close to the sun as well as long-duration orbits will experience a cumulative perturbation.

\section*{Acknowledgments}

We would like to thank Gregory Matloff for useful conversations.


\end{document}